\documentclass[prd,twocolumn,preprintnumbers,nofootinbib]{revtex4}

\usepackage{graphicx}
\usepackage{dcolumn}
\usepackage{bm}
\usepackage{amsfonts}
\usepackage{amsmath}
\usepackage{amsthm}
\usepackage{amssymb}

\newcommand{\C}[1]{{\mathcal #1}}

\newcommand{\beq}{\begin{equation}}
\newcommand{\eeq}{\end{equation}}
\newcommand{\bea}{\begin{eqnarray}}
\newcommand{\eea}{\end{eqnarray}}
\newcommand{\nn}{\nonumber}

\newcommand{\half}{\frac 12}

\begin{document}

\date{\today}
\title{Preheating after N-flation}

\author{Diana Battefeld} \email[email: ]{diana.battefeld(AT)helsinki.fi}
\affiliation{Helsinki Institute of Physics, P.O. Box 64, University of Helsinki, FIN-00014 Helsinki, Finland}

 \author{Shinsuke Kawai} \email[email: ]{shinsuke.kawai(AT)helsinki.fi}
 \affiliation{Helsinki Institute of Physics, P.O. Box 64, University of Helsinki, FIN-00014 Helsinki, Finland}

\preprint{HIP-2008-06/TH}
\begin{abstract}
We study preheating in ${\mathcal N}$-flation, assuming the Mar\v cenko-Pastur mass distribution, 
equal energy initial conditions at the beginning of inflation and equal axion-matter couplings, 
where matter is taken to be a single, massless bosonic field. 
By numerical analysis we find that preheating via parametric resonance is suppressed, 
indicating that the old theory of perturbative preheating is applicable. 
While the tensor-to-scalar ratio, the non-Gaussianity parameters and the scalar spectral index 
computed for $\C N$-flation are similar to those in single field inflation  (at least within an
observationally viable parameter region), our results suggest that the physics of preheating can 
differ significantly from the single field case.
\end{abstract}
\maketitle
\newpage


\section{Introduction}

While inflationary cosmology is becoming a precision science, with the advent of recent and upcoming experiments such as the measurement of the CMBR by the Planck satellite \cite{:2006uk},
the particle physics origin of inflatons still remains unclear.
Due to their simplicity, single field models of inflation are considered the most economical 
explanation of a Gaussian, nearly scale invariant spectrum of primordial fluctuations, as well as the flatness and the large-scale homogeneity of the observed universe.
Future experiments, however, could change this situation and put single field models under pressure.
For these reasons, there has been a keen interest in building multi-field inflationary models, see e.g. the reviews \cite{Wands:2007bd,Bassett:2005xm}.
Promising setups of multi-field inflation include string-motivated models such as $\mathcal{N}$-flation 
\cite{Dimopoulos:2005ac}, inflation from multiple M5-branes \cite{Becker:2005sg} and inflation from tachyons \cite{Majumdar:2003kd}.

Naturally, the large parameter space for couplings, masses and initial conditions pertaining to multi-field inflation makes a systematic analysis difficult.
An interesting exception, nevertheless,  is $\mathcal{N}$-flation.
$\mathcal{N}$-flation is a string motivated implementation of assisted inflation 
\cite{Kanti:1999vt,Kanti:1999ie,Liddle:1998jc,Copeland:1999cs,Singh:2006yy,Kaloper:1999gm} 
where a large number of uncoupled scalar fields, identified with axions arising from KKLT compactification of type IIB string theory, assist each other to drive an inflationary phase
\cite{Dimopoulos:2005ac}; see also 
\cite{Easther:2005zr,Kim:2006ys,Kim:2006te,Piao:2006nm,Grimm:2007hs}. 
One salient feature of this model is the possible avoidance of super-Planckian initial values. 
Further, using results from random matrix theory, the masses for the axion fields can be shown to conform to the Mar\v cenko-Pastur (MP) law \cite{Easther:2005zr} under  reasonable approximations.
The spectrum of the masses is controlled by only two parameters, the average mass and a variable controlling  the ratio between the number of axions and the total dimension of the moduli space.  
This renders $\mathcal{N}$-flation  tractable despite the large number of fields. 
To a certain extent, successful inflation is contingent upon  the  initial conditions, however, the model becomes easily tractable by assuming that  each field possesses equal initial energy.

Observable cosmological imprints from $\mathcal{N}$-flation have been computed by several groups.
The tensor-to-scalar ratio $r$ was calculated by Alabidi and Lyth \cite{Alabidi:2005qi} and was shown to 
have the same value as in the single field case.
The non-Gaussianity parameter $f_{NL}$ was computed in \cite{Kim:2006te,Battefeld:2007en} where the deviation from  single field models was found to be negligible.
Unlike $r$ and $f_{NL}$,  the spectral index of the curvature perturbation $n_s$ depends  slightly on the model's parameters.
\cite{Easther:2005zr,Kim:2006te,Piao:2006nm} showed  that $n_s$ is smaller (the spectrum being redder) than that found in  single field models, in agreement with the general discussion made in
\cite{Lyth:1998xn}.
These results suggest that the observational data predicted by $\mathcal{N}$-flation are not drastically different from the single-inflaton case.
Note however that the 5-year WMAP data already excludes some region of the parameter space; 
see \cite{Komatsu:2008hk}.

In this article, we study  multi-field preheating, focusing on  $\mathcal{N}$-flation as a specific example. 
To our knowledge, a general theory of preheating for multi-field inflationary models has not been fully developed. 
This is in part due to the highly non-trivial nature of the string theoretical constructions responsible for inflation. However,  even at the phenomenological level, effects due to multiple inflatons contributing to preheating are largely unexplored in contrast to single field inflation 
\cite{Dolgov:1982th,Dolgov:1989us,Traschen:1990sw,Kofman:1997yn,Shtanov:1994ce}, see also 
\cite{Khlebnikov:1996mc,Khlebnikov:1996wr,Khlebnikov:1996zt,Prokopec:1996rr,Podolsky:2005bw,Dufaux:2006ee}. 
For instance, preheating via parametric resonance of a matter field might be more efficient in the 
presence of multiple inflatons, as indicated in Cantor preheating \cite{Bassett:1997gb, Bassett:2005xm}. Here, a non-periodic variation of the matter field's effective mass leads to the dissolution of the stability bands and a possible parametric amplification of almost all Fourier modes. 
This expectation is based on spectral theory \cite{JMoser,gihman,kirsch,Pastur},
but a quantitative study about the magnitude of the amplification  with more than two fields is missing \cite{Bassett:1998yd}.
Owing to  the possible dissolution of the stability bands, one might  expect  that the
collective behavior of the fields  gives rise to efficient particle production after 
$\mathcal{N}$-flation.

To introduce the notation, let us consider a single field model first: 
the equation of motion for a matter field $\chi_k$ with wavenumber  $k$ 
(assuming the coupling $g^2\varphi^2\chi^2/2$ )  can be written as a  Mathieu equation \cite{Kofman:1997yn},
\begin{equation}
\frac{d^2 X_k}{d\tau^2}+\left[ A-2q\cos(2\tau)\right] X_k=0,
\end{equation}
for the comoving matter field $X_k=a^{3/2}\chi_k$, where
$\tau=m_\varphi t$ is the rescaled time, $m_\varphi$ the inflaton mass, and the two resonance parameters are
\begin{equation}
q=\frac{g^2\Phi_0^2}{4m_\varphi^2}, ~~~~~
A=\frac{k^2}{a^2m_\varphi^2}+2q.
\end{equation}
Here, $\Phi_0$ is the slowly varying inflaton amplitude, $g$ is the inflaton-matter coupling and 
$a$ the scale factor (see also section \ref{icp}). 
The efficiency of  parametric resonance is controlled by the resonance parameter $q$, which needs to be large enough ($q\gg 1$) in order for the resonance effect to hold against  cosmic expansion.
The upper bound on the coupling constant $g$ is given by the  potential's stability condition  
against quantum gravity effects as well as  radiative corrections \cite{Zlatev:1997vd,Kofman:1997yn}
(unless the potential is protected by supersymmetry).
This upper bound on $g$ and $m_\varphi\sim 10^{-6}M_{P}$ ($M_{P}$ is the reduced Planck mass) from the COBE normalisation restricts the aforementioned  resonance parameter $q$, leaving not much room for effective parametric resonance \cite{Zlatev:1997vd}. One might hope to alleviate this fine-tuning in multi-field models.

For definiteness, we focus  on  $\mathcal{N}$-flation.
We  use the Mar\v cenko-Pastur mass distribution for the axion masses, put forth  
in \cite{Easther:2005zr} based on random matrix theory, choosing  the most likely mass distribution, 
see section \ref{sec:1}. 
Further, for simplicity,  we assume  equal energy initial conditions at the onset of inflation. 
Recently, aspects of preheating in the context of $\mathcal{N}$-flation have been considered in \cite{Green:2007gs}, pointing out the danger of transferring energy preferably to hidden sectors instead of standard model particles. This reveals an additional need for fine tuning, a possible problem for many string-motivated models of inflation.  The study in \cite{Green:2007gs} is based entirely on an effective single field description of $\mathcal{N}$-flation, such as the one above.  
Thus the common lore of parameteric resonance models seems to be applicable in this work. 
Here we take the optimistic view that preheating might indeed occur in the visible sector. 
However, we go beyond the single field model. 
We find that at the end of inflation more than 90 percent of the energy is confined to the lightest 10
percent of the fields in a very narrow mass range, while the remaining heavier fields are already more or less settled to the bottom of their potentials. 
Thus we focus on these lightest $\tilde{\mathcal{N}}$ 
 fields during preheating.
We also show that  slow roll is indeed a good approximation, even though heavier fields violate the slow roll condition $|\eta|<1$ long before preheating starts. 
The initial values for preheating of the crucial $\tilde{\mathcal{N}}$ light fields are therefore given by their slow roll values at the end of inflation.
Since the lightest fields are highest up in their potentials, fields will join preheating in a staggered fashion. This and the still reasonably large number of fields makes a numerical treatment mandatory, which we provide.
We solve the equations of motion for the matter field for various parameters following the above prescription and find that  parametric resonance is less effective for ${\C N}$-flation.  The physical reason for the suppression of parametric resonance  is the dephasing of the multiple fields. It is thus clear that effective single field models fail to properly account for this effect. We conclude that  the old theory of perturbative preheating,  and not parametric resonance, is applicable when many fields 
of different masses couple to a single matter field.

This article is structured as follows: in section \ref{sec:1} we review $\mathcal{N}$-flation and its dynamics during slow roll. 
We extend this discussion in section \ref{sec:bsr}, where we provide an extrapolated slow roll solution
that slightly underestimates the inflatons' potential energy during inflation. 
We compare the slow-roll solution with a numerical solution and argue that even after $\eta<1$ is violated by one or more of the heavier axion fields, the overall behavior of the inflatons is still well
approximated by the slow-roll regime, up until preheating commences.
In section \ref{icp}  we set the stage for preheating corresponding to the end of slow roll for the effective single field model.  To begin, we set  all axion masses equal  to each other and discuss the physics of preheating in this particular system;  then
we return to  ${\C N}$-flation, where the Mar\v cenko-Pastur mass spectrum is used.
Finally, in section \ref{sec:conc} we conclude with comments and prospects for studying further issues on multi-field preheating.
In Appendix \ref{AppA} we give a semi-analytic solution that provides an upper bound for the inflaton
potential, which further supports the observation made in section \ref{sec:bsr}.

\section{$\mathcal{N}$-flation and Slow Roll \label{sec:1}}

The action for $\mathcal{N}$ minimally coupled scalar fields, responsible for driving an
inflationary phase, can be written as (see \cite{Wands:2007bd} for a review on multi-field
inflation)
\begin{equation}
S\!=\!-\!\int\!\! d^4x \sqrt{-g}\! \left(\!\frac{1}{2}
\sum_{i=\!1}^{\mathcal{N}}g^{\mu\nu}\nabla_\mu\varphi_i\nabla_\nu\varphi_i
\!+\!W(\varphi_1,\varphi_2,...\!)\!\right)\,,
\end{equation}
where we assume canonical kinetic terms.
The unperturbed volume expansion rate from an initial hypersurface at $t_*$ to a final 
hypersurface at $t_c$
(below we use $*$ and $c$ to denote values evaluated at $t_*$ and $t_c$)
is given by
\begin{eqnarray}
N(t_c,t_*)\equiv\int_{*}^{c} H dt \,,\label{nofh}
\end{eqnarray}
where $N$ is the number of e-folds, $H$ is the Hubble parameter and $t$ is cosmic time.

In $\mathcal{N}$-flation \cite{Dimopoulos:2005ac}, the $\mathcal{ N}$
scalar fields that drive inflation are associated  with axion fields. All
cross-couplings vanish when the periodic potentials are expanded
around their minima \cite{Easther:2005zr}.
Therefore, in the proximity of their minima
the fields have a potential of the form
\begin{eqnarray}
W(\varphi_1,\varphi_2,\dots,\varphi_{\mathcal{N}})&=&\sum_{i=1}^{\mathcal{N}}V_i(\varphi_i)\nn\\
&=&\sum_{i=1}^{\mathcal{N}}\frac{1}{2}m_i^2\varphi_i^2\,,\label{potential}
\end{eqnarray}
where  the fields have been arranged according to the magnitude of their masses,
namely $m_i>m_j$ if $i>j$.
$\mathcal{N}$-flation is a specific realization\footnote{
For a different realization of assisted inflation based on M-theory from
multiple M5-branes see \cite{Becker:2005sg} \cite{Krause:2007jr}.
See also \cite{Podolsky:2007vg} for another approach to random potentials in the landscape.
} 
of assisted inflation
\cite{Liddle:1998jc,Kanti:1999vt,Malik:1998gy},
where the $\mathcal{N}$  scalar fields assist each other in driving an
inflationary phase. 
In this manner, individual  fields do not need to traverse a super-Planckian stretch in field space.
The spectrum of masses in (\ref{potential}), which were assumed to be  equal 
in \cite{Dimopoulos:2005ac}, can be evaluated by means of random matrix theory within a context of
KKLT moduli stabilisation \cite{Kachru:2003aw}, and was found by Easther and McAllister to conform to the Mar\v{c}enko-Pastur (MP) law \cite{Easther:2005zr}. 
This results in a probability for a given square mass of
\begin{eqnarray}
p(m^2)&=&\frac{1}{2\pi m^2\beta \bar{m}^2}\sqrt{(m_{max}^2-m^2)(m^2-m_{min}^2)}\,, \label{mpdistr}
\end{eqnarray}
where $\beta$ and $\bar{m}^2$ completely describe the distribution. 
Here, $\bar{m}^2$ is the average mass squared and $\beta$ controls the width and shape of the spectrum. 
The smallest and largest masses are given by
\begin{eqnarray}
m_{min}&
\equiv&\bar{m}(1-\sqrt{\beta})\,,
\label{mmin}\\
m_{max}&
\equiv&\bar{m}(1+\sqrt{\beta})\,.
\end{eqnarray}
In this paper we split the mass range $(m_{min}, m_{max})$ into ${\C N}$ bins,
\begin{eqnarray}
(\tilde m_0, \tilde m_1), 
(\tilde m_1, \tilde m_2), \cdots, 
(\tilde m_{{\C N}-1}, \tilde m_{\C N})
\label{Nbins}
\end{eqnarray}
where $\tilde m_0=m_{min}$, 
$\tilde m_{{\C N}}=m_{max}$ and $\tilde m_{i-1}<\tilde m_{i}$,
so that
\begin{eqnarray}
\int_{\tilde m_{i-1}^2}^{\tilde m_{i}^2}p(m^2)dm^2=\frac{1}{\C N},~~~~~
i=1, 2, \cdots, {\C N}.
\label{eqn:EqualSplit}
\end{eqnarray}
We then represent each bin $(\tilde m_{i-1}, \tilde m_{i})$ by an inflaton of mass $m_i$.
In practice we simply set 
\begin{eqnarray}
m_i^2=(\tilde m_{i-1}^2+\tilde m_i^2)/2,~~~~~
i=1, 2, \cdots, {\C N},
\label{eqn:MidPointMass}
\end{eqnarray}
in  the numerical computations.
Apart from the ${\C N}$ inflatons $\varphi_1,\varphi_2,\cdots,\varphi_{\C N}$ with masses
$m_1,m_2,\cdots,m_{\C N}$, we introduce a fiducial inflaton $\varphi_{0}$ with mass
$m_{0}=m_{min}$ for computational convenience (we shall use this as a clock).
In \cite{Easther:2005zr}, $\beta$ is identified with the number of
axions contributing to inflation divided by the total dimension of the
moduli space (K\"ahler, complex structure and dilaton) in a given KKLT
compactification of type IIB string theory. 
Due to constraints arising from the
renormalization of Newton's constant \cite{Dimopoulos:2005ac}  $\beta\sim 1/2$ is preferred. 
Hence, we will work with $\beta=1/2$ in the following. 
Further, the magnitude of $\bar{m}$ is constrained by the COBE normalization 
\cite{Easther:2005zr,Gong:2006zp}, so that there is not much freedom in 
$\mathcal{N}$-flation to tune parameters.

\begin{figure}[tb]
  \includegraphics[scale=0.8]{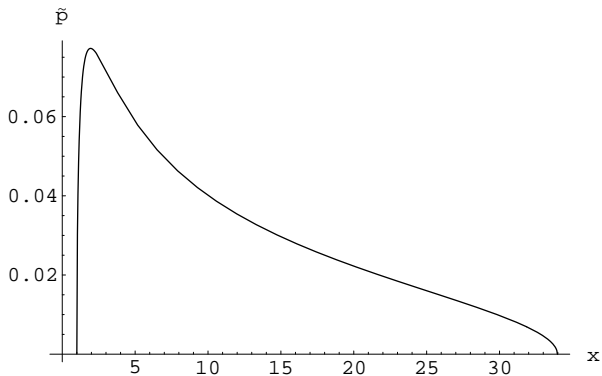}
   \caption{A plot of the Mar\v cenko-Pastur mass distribution over the dimensionless mass 
   variable $x=m^2/m_{min}^2$, for $\beta=0.5$.
   Note that the MP distribution peaks in the small-mass region so that the majority of fields resides 
   there.
   We use mass parameters $x_i=m_i^2/m_{min}^2$ corresponding to the mass-bins chosen according
   to (\ref{Nbins}), (\ref{eqn:EqualSplit}), (\ref{eqn:MidPointMass}), so that 
   $x_0\equiv 1<x_1<x_2<\cdots<x_{{\C N}-1}<x_{\C N}<\xi\approx 34$.
\label{fig0}}
\end{figure}

At this point, we introduce a convenient dimensionless mass parameter
\begin{eqnarray}
x_i&\equiv&\frac{m_i^2}{m_{min}^2}\,,
\label{eqn:defx}
\end{eqnarray}
as well as the suitable short-hand notation
\begin{eqnarray}
\xi\equiv\frac{m_{max}^2}{m_{min}^2}=
\left(\frac{1+\sqrt\beta}{1-\sqrt\beta}\right)^2\,.
\label{xi}
\end{eqnarray}
A properly normalised probability distribution for  the variable $x=m^2/m_{min}^2$ is 
$
\tilde p(x)=m_{min}^2p(m^2);
$
hence expectation values with respect to the MP-distribution can be evaluated via
\begin{eqnarray}
\hspace{-10mm}
\left<f(x)\right>&
\equiv& \frac{1}{\mathcal{N}} \sum_{i=1}^{\mathcal{N}}f(x_i) \label{defexp0}\nn\\
&=&\int_1^\xi\tilde p(x) f(x)\, dx\nn\\
&=&\frac{(1-\sqrt\beta)^2}{2\pi\beta}\int_1^\xi\sqrt{(\xi-x)(x-1)}\frac{f(x)}{x}\,dx
\label{MPintegral}\,.
\end{eqnarray}
In section \ref{sec:lad} and in appendix \ref{AppA}, we  make use of the additional notation
\begin{eqnarray}
\left<f(x)\right>\Big|_a^b\equiv\frac{(1-\sqrt\beta)^2}{2\pi\beta}
 \int_a^b\sqrt{(\xi-x)(x-1)}\frac{f(x)}{x}\,dx\label{MPboundary}\,,
\end{eqnarray}
where $a$ and $b$ are the limits of integration. 
When $f(x)$ is a polynomial in $x$, (\ref{MPintegral}) reduces to a hypergeometric integral.
In particular,
\begin{eqnarray}
\left<x^{-1}\right>=\xi^{-1/2}, ~~~
\left<1\right>=1,~~~
\left<x\right>=\frac{1}{(1-\sqrt{\beta})^2}.
\end{eqnarray}
Fig.\ref{fig0} shows a plot of the probability distribution function $\tilde p(x)$, for $\beta=1/2$.

Initially, we restrict ourselves to the slow roll approximation. 
As we have already seen, $\mathcal{N}$ fields contribute to the
energy density of the universe through a separable potential. In this
regime, the dynamics of  $\mathcal{N}$-flation is as follows:  first note that the field equations and Friedmann equations can be written as
\begin{eqnarray}
3H\dot{\varphi}_i &\approx&-\frac{\partial V_i}{\partial \varphi_i}\equiv- V_i^{\prime}\,,\\ \label{friedman}
3H^2&\approx& W \label{klein}\,.
\end{eqnarray}
Here and throughout most of our analysis, we set the reduced Planck mass to 
$M_P=(8\pi G)^{-1/2}\equiv 1$. 
The  slow roll approximation is valid if the parameters
\begin{eqnarray}
\varepsilon_i\equiv\frac{1}{2}\frac{V_{i}^{\prime 2}}{W^2}\hspace{0.5cm},\hspace{0.5cm}\eta_i\equiv\frac{V_{i}^{\prime\prime}}{W}\,, \label{srparameters}
\end{eqnarray}
are small ($\varepsilon_i\ll 1$, $\eta_i\ll 1$) and
\begin{eqnarray}
\varepsilon\equiv\sum_{i=1}^{\mathcal N}\varepsilon_i\ll1
\label{eqn:epsilon}
\end{eqnarray}
holds. The number of e-folds of inflation becomes
\begin{eqnarray}
N(t_c,t_*)=-\sum_{i=1}^{\mathcal N}\int_*^c \frac{V_i}{V_i^\prime} d\varphi_i\,,\label{defN}
\end{eqnarray}
and the field equations can be integrated to yield
\begin{eqnarray}
\left(\frac{\varphi_i^c}{\varphi_i^*}\right)^{{1}/{m_i^2}}
=\left(\frac{\varphi_j^c}{\varphi_j^*}\right)^{{1}/{m_j^2}}\,. \label{solphi}
\end{eqnarray}
Notice that this relationship between fields does not correspond to an
attractor solution and predictions of $\mathcal{N}$-flation can
depend on the initial conditions.
Recall that  we are assuming equal energy initial conditions, namely
$V_i^*=V_j^*$, which can be rewritten as
\begin{equation}
m_i\phi_i^*=m_j\phi_j^*\,. \label{equalenergy}
\end{equation}
A subtle feature of $\mathcal{N}$-flation is that if the mass
spectrum is broad ($\xi\gg 1$, corresponding to $\beta\sim 1$), 
the heavier fields will acquire $\eta_i \approx 1$, 
even as  inflation continues. 
This is the case even for the preferred value of $\beta\sim 1/2$, 
corresponding to $\xi\sim 34$. 
In the next section we  show how to deal with
$\mathcal{N}$-flation after $\eta_i\sim1$ for the heaviest inflaton.

\section{Evolution of inflatons and an effective single field model during Inflation 
\label{sec:bsr}}

In this section we investigate the evolution of the axion fields after the slow 
roll condition is violated for one or more of them.
In order to study the system's collective behavior, it is useful to use
an effective single field description \cite{Wands:2007bd}.
Given equal energy initial conditions for the fields, the slow roll parameter 
$\eta_{\mathcal{N}}$ of the heaviest field will be the first one to become of order unity
\cite{Battefeld:2006sz}. 
Hence, prior to this moment we can safely implement an effective model composed
of a single field $\sigma$ which evolves according to an effective potential 
$W_{eff}(\sigma)$. 
After $\eta_{\mathcal{N}}$ became of order one, the corresponding field 
$\varphi_{\mathcal{N}}$ cannot be described by the slow-roll solution. 
Below we argue that in our particular model of ${\C N}$-flation the whole system
is nevertheless well approximated by the slow-roll solution, up until the slow-roll parameter
$\varepsilon$ of (\ref{epsilon}) becomes of order one\footnote{Or the slow roll parameter of the effective degree of freedom in (\ref{epsilon}).}; during this stage, the contribution 
of the heavy axions is negligible compared to that of the lighter fields.
This is due to three characteristic features of the model: 
(1) the  majority of the axions is distributed around the lightest mass in the MP law;
(2) the small value of the heaviest fields is prescribed by the equal energy initial condition; hence, heavy fields provide a
small contribution from the onset;
(3) when the slow-roll condition $\eta_{i}<1$ is violated for  the heaviest fields the
Hubble parameter is still very large, resulting in an over-damped evolution of the heavy fields.

Naturally, it is possible to continue using the slow roll approximation when the next  heaviest 
field violates slow roll and so on and so forth.
This regime ends when  slow roll fails  for the effective single field $\sigma$, 
after which light fields will actually start to evolve faster and preheating starts.
It is important to note  that we can trust our approximation up until preheating starts, where possible particle production due to non-linear parametric resonance is our main concern.

\subsection{Effective single field model\label{sec:A}}

Here  we derive the effective single-field model based on the slow roll
approximation.
This provides a lower bound to the evolution of the total potential energy. 
We identify the effective inflaton field $\sigma$ as the path-length of the trajectory in the $\mathcal{N}$ dimensional field space; namely, for $\mathcal{N}$ scalar fields $\phi_i$ we have \cite{Wands:2007bd}
\begin{eqnarray}
\sigma\equiv\int_{t_*}^{t}\sum_{i=1}^{\mathcal{N}}{\hat{\sigma}_i\dot{\varphi_i}dt}\,,\label{defsigma}
\end{eqnarray}
with
\begin{eqnarray}
\hat{\sigma}_i\equiv\frac{\dot{\varphi_i}}{\sqrt{\sum_j{\dot{\varphi}_j^2}}}\,, \label{defsigma2}
\end{eqnarray}
where the $\varphi_i$ can be computed given the dynamical relations in (\ref{solphi}), which are valid during slow roll, as well as the initial conditions in (\ref{equalenergy}). 
Note that $\sigma=0$ at the initial time $t_*$.

Using
\begin{eqnarray}
y\equiv\frac{\varphi_{0}^{2}}{\varphi_{0}^{*2}}
\end{eqnarray}
(where $y$ parameterizes how far the field $\varphi_0$ rolls down its potential) and $x_i$  is defined in
(\ref{eqn:defx}), 
as well as the  equal energy initial conditions (\ref{equalenergy}),
we can rewrite the dynamical relations as
\begin{eqnarray}
\varphi_i^2=\varphi_{0}^{*2}\frac{y^{x_i}}{x_i}\,.
\end{eqnarray}
We are using the fiducial inflaton $\varphi_{0}$ for computational convenience; $\varphi_{0}$
is not one of  the ${\C N}$ inflatons driving ${\mathcal N}$-flation 
(note that there is a  vanishing probability for $m=m_0=m_{min}$ according to the MP law). 
From the Klein-Gordon equations during slow roll along  with the Friedmann equation we obtain
$\dot{\varphi}_i^2=m_{0}^4x_i\varphi_{0}^{*2}y^{x_i}/(3W)$, as well as 
$dy=-(2m_{0}^2y/\sqrt{3W})dt$,  with $W=(1/2)m_{0}^2\varphi_{0}^{*2}\sum_{i=1}^{\C N}y^{x_i}$.
Using these relations in (\ref{defsigma}), we obtain an effective single-field solution (for which the subscript $I$ is used), 
\begin{eqnarray}
\sigma_I(y)&=&
\frac{-\varphi_{0}^{*}}{2}\int_1^y{{\left(\sum_{i=1}^{\mathcal{N}}x_is^{x_i}\right)^{1/2}}\frac{ds}{s}}\nn\\
&=&\frac{\sqrt{\mathcal{N}}}{2}\varphi_{0}^*\int_y^{{1}}\sqrt{\left<x s^x\right>}\frac{ds}{s}\,,\label{sigma1}
\end{eqnarray}
where in the last step we used the definition of the MP-expectation values from (\ref{defexp0}). 
Similarly, the corresponding potential in (\ref{potential}) can be computed as
\begin{eqnarray}
W_I(y)
&=&\half m_{0}^2\varphi_{0}^{*2}\sum_{i=1}^{\C N}{y^{x_i}}\nn\\
&=&\frac{\mathcal{N}}{2}m_{0}^2\varphi_{0}^{*2}\left<y^x\right>\,. \label{W1}
\end{eqnarray}
Equations (\ref{sigma1}) and (\ref{W1}) provide an implicit means of computing $W_I(\sigma_I)$.
The number of  e-folds can also be computed with this approximation. 
From (\ref{defN}) we find,
\begin{eqnarray}
N_I(y)
&=&\frac{1}{4}\varphi_{0}^{*2}\sum_{i=1}^{\mathcal{N}}\Big(\frac{1}{x_i}-\frac{y^{x_i}}{x_i}\Big)\,\nn\\
&=&\frac{\mathcal{N}}{4}\varphi_{0}^{*2}\left[\left<x^{-1}\right>-\left<x^{-1}y^x\right>\right]\,.\label{efold1}
\end{eqnarray}

Let us look at the solution more quantitatively. 
Since we would like to avoid super-Planckian initial conditions we assume the lightest possible inflaton to set off from the reduced Planck scale,
$\varphi_{0}^*= 1$.
Then it follows from the equal energy initial conditions that all the other fields evolve safely within sub-Planckian scale.
Here and in the following we use  the preferred $\beta=1/2$ so that the ratio of the heaviest to the 
lightest mass squared in (\ref{xi}) is about $\xi\approx 34$.
The e-folding number (\ref{efold1}) depends linearly on the number of inflatons ${\C N}$.
If we take $\mathcal{N}=1500$ we get $N_{max}\equiv N_I(0)\approx 64.3$, which is large enough to solve the standard cosmological problems.
Note, however, that one cannot trust (\ref{sigma1}) and (\ref{W1}) down to $y=0$, since slow roll ends earlier. 
Strictly speaking, our effective single field solution (with subscripts `$I$') is only valid as long as 
the slow roll conditions are satisfied, that is until
$\eta_{\mathcal{N}}=1$.
Using (\ref{W1}) and $m_{\C N}\approx m_{max}$ this can be written as
\begin{eqnarray}
\left<y^x\right>=\frac{2\xi}{\mathcal{N}\varphi_{0}^{*2}}\,, \label{yN}
\end{eqnarray}
from which the value of $y$ at $\eta_{\C N}=1$ is found numerically as
$y_{\C N}\approx 0.488$ for ${\C N}=1500$. 
The number of e-folds at this instant is $N_I(y_{\C N})\approx 55.6$ and we see that there is still a breadth of inflation to come. 
If we ignore this fact, we could extrapolate $\sigma_I$ up until this effective degree of freedom 
leaves its own slow rolling regime when
\begin{eqnarray}
\epsilon_{\sigma}\equiv\frac{1}{2}\left(\frac{W_I^\prime}{W_I}\right)^2=1\,.\label{epsilon}
\end{eqnarray}
This equation can be rewritten as
\begin{eqnarray}
2\left<xy^x\right>=\mathcal{N}\varphi_{0}^{*2}\left<y^x\right>^2\,, \label{yend}
\end{eqnarray}
where we used
\begin{eqnarray}
W_I^\prime\equiv\frac{\partial W_I}{\partial \sigma_I}&=&\sum_{i=1}^{\mathcal{N}}\hat{\sigma_i}\frac{\partial V_i}{\partial \varphi_i}\nn\\
&=&\sum_{i=1}^{\mathcal{N}}\hat{\sigma_i}m_i^2\varphi_i\nn\\
&=&m_{0}^2\varphi_{0}^{*}\sqrt{\mathcal{N}\left<xy^x\right>}\,.
\end{eqnarray}
Equation (\ref{yend}) can be numerically solved to obtain
 $y_{end}\approx 0.0836$ for ${\C N}=1500$ so that
$\sigma_I(y_{end})\approx 17.6$ and $N_I(y_{end})\approx63.8$.
At this point inflation comes to an end and preheating is about to commence. 
The potential at this instant is $W_I(y_{end})\approx 0.128 \bar{m}^2\approx 1.49 m_0^2$ 
(see Table \ref{table1}).

\begin{figure}[tb]
  \includegraphics[scale=0.8]{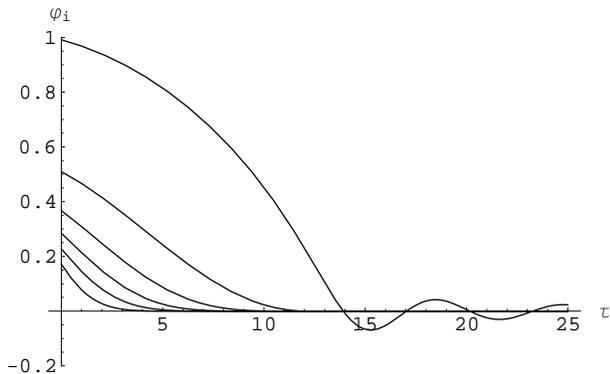}
   \caption{
   Evolution of the axion fields for ${\C N}=1500$ (numerical result). 
   The figure shows $\varphi_i$ with $i=1,300,600,900,1200,1500$ from the top.
   The time $\tau=m_0 t$ is in units measured by the fiducial mass $m_{0}=m_{min}$.
   In this unit, $y=y_{\C N}$ is at $\tau=7.14$ and $y=y_{end}$ is at $\tau=11.6$.
    \label{fig2}}
\end{figure}
\subsection{Implications for non-Gaussianities}

In contrast to single-field inflationary models, in which non-Gaussianities (NG) are known to be suppressed,
in multi-field models there is a possibility that NG may become large due to the existence of 
isocurvature perturbations \cite{Gordon:2000hv};
if there is a sudden turn of the trajectory in field space as in the case of the curvaton scenario, a conversion of isocurvature perturbations into adiabatic ones takes place, giving rise to larger NG.
In $\mathcal N$-flation the non-linearity parameters characterizing the bi- and tri-spectrum were investigated in the horizon crossing approximations in \cite{Battefeld:2007en} and it was found that 
$\mathcal N$-flation is indistinguishable from single field inflationary models in this limit. 
Also, incorporating the evolution of perturbations after horizon crossing, but still within slow roll, revealed that additional contributions remain negligible \cite{Battefeld:2007en}. 
Hence,  NG are expected to be heavily suppressed as long as slow roll is considered
\cite{Battefeld:2006sz}.
In the present paper we investigate the evolution in ${\mathcal N}$-flation after the slow roll condition is violated for one or more of the axion fields and find (see the next sections) that the extrapolated slow-roll solution remains a good approximation up until preheating commences.
Thus, in this intermediate regime (after slow roll inflation but before preheating), NG should also be suppressed; additional NG would be due to the
evolution of the adiabatic mode after horizon crossing; for this to occur, isocurvature modes have to source the adiabatic one,
but in $\mathcal{N}$-flation, the trajectory in field space is smooth; therefore, NG should be heavily suppressed up until slow roll fails for the effective single degree of freedom, i.e. when
preheating begins.
During preheating NG may still appear; this requires further study, and we hope to come back to this issue in the near future.

\subsection{Numerical solutions\label{sec:disc1}}

Fig.\ref{fig2} shows the  time evolution of ${\C N}=1500$ axions in our setup, 
namely the MP mass distribution with $\beta=0.5$ and the equal energy initial conditions,
obtained numerically. 
The plot shows  $\varphi_i$ with $i=1,300,600,900,1200$ and $1500$.
For the initial conditions we used (\ref{equalenergy}) with $\varphi_0^*=1$ and $\dot\varphi^*_i=0$.
Due to these initial values, lighter axions evolve from larger values, closer to $1$.
The figure clearly shows that heavy axions, even $\varphi_{300}$, roll down the potential rapidly  and their oscillation amplitudes are much smaller than that of the lightest field $\varphi_1$. Indeed, fields are usually over-damped up until preheating starts.
Naturally, the lighter fields are expected to be responsible for  preheating 
(unless the coupling between the heavy fields and a matter field is extremely strong).

In Table \ref{table1} we summarise the values of $W$ and $N$ obtained by the extrapolated
slow-roll solutions ($I$) and numerical results, at $\sigma=\sigma(y_{\C N})$ and $\sigma(y_{end})$. 
A comparison between the analytic and numerical computation for  ${\C N}=100, 200, 400,1500$ reveals
an agreement,  roughly within $15\%$, indicating that the extrapolated slow-roll solution ($I$) is a good 
approximation up until $y_{end}$. 
In Appendix \ref{AppA} we provide a semi-analytic computation of an upper bound to $W$ 
(solution $II$),
which further supports this observation.
How many inflatons still satisfy the slow-roll condition $\eta_i<1$ at $y_{end}?$
This can be found by comparing the values of $W(y_{end})/m_{0}^2$ and the mass parameter
$x_i$, since
$\eta_i=m_i^2/W=x_im_{0}^2/W$.
Numerically, we find  that 
$13$, $18$, $25$, $56$ lightest fields are still in the slow-roll regime at $y=y_{end}$, 
for ${\C N}=100, 200, 400, 1500$. 
Analytically,  solution $I$ yields somewhat smaller values:
$10$, $13$, $18$, $41$, respectively.

\begin{table}[htdp]
\begin{center}\begin{tabular}{c||cc|cc|cc}
&&& \multicolumn{2}{c|}{solution I} & \multicolumn{2}{c}{Numerical}\\
\cline{4-7}
${\C N}$& $y_{\C N}$ & $\sigma_I(y_{\C N})$ &
$W_I(y_{\C N})$ & $N_I(y_{\C N})$ & 
$W(y_{\C N})$ & $N(y_{\C N})$\\
 & $y_{end}$ & $\sigma_I(y_{end})$ & $W_I(y_{end})$ & $N_I(y_{end})$ & 
$W(y_{end})$ & $N(y_{end})$\\
\hline
100 & 0.964 & 0.541 & 34.0 & 0.758 & 34.0 & 1.10\\
 & 0.502 & 3.27 & 2.43 & 3.67 & 2.83 & 4.34\\
\hline
200 & 0.879 & 2.00 & 34.0 & 3.72 & 34.3 & 4.14\\
 & 0.331  & 5.40 & 1.97 & 8.00 & 2.27 & 8.83\\
\hline
400 & 0.762 & 4.37 & 34.0 & 10.9 & 34.4 & 11.4\\
 & 0.211 & 8.33& 1.73 & 16.6 & 1.95 & 17.6\\
\hline
1500 & 0.488 & 12.86& 34.0 & 55.6& 34.4 & 56.3 \\
 & 0.0836 & 17.61& 1.49 & 63.8 & 1.62 & 65.1
\end{tabular} 
\caption{Comparison of analytic and numerical solutions for the effective
single-field values $W$ and $N$ at $\sigma(y_{\C N})$ and $\sigma(y_{end})$, 
for the number of inflatons ${\C N}=100, 200, 400$, and $1500$.
The values of $\sigma_I$ are found using (\ref{sigma1}) and the corresponding
analytic and numerical values for $W/m_{0}^2$ and $N$ are shown.
Apart from the conspicuous disagreement in the e-folding number $N$ for small
${\C N}$, the extrapolated slow-roll solutions (I) are relatively in good agreement with 
the numerical solutions, roughly within $15\%$ difference.
Typically,  the results of solution (I) slightly underestimate the potential $W$.
\label{table1}}
\end{center}
\end{table}

\subsection{Light axion dominance\label{sec:lad}}

\begin{figure}[tb]
  \includegraphics[scale=0.8]{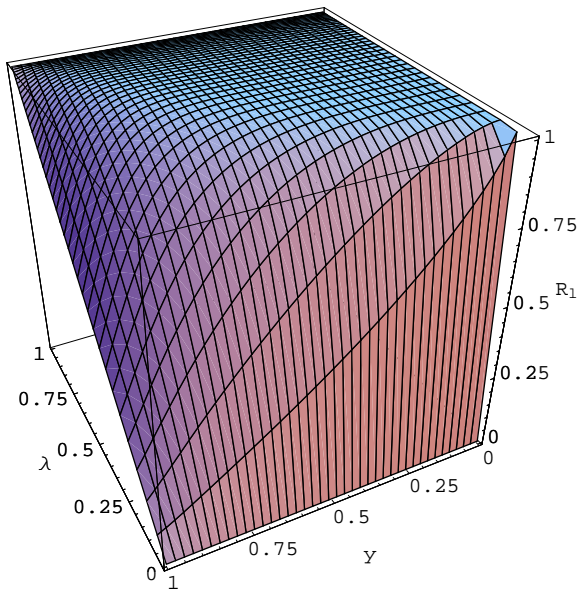}
  \caption{
  The ratio of potential energies of the light fields to all  axions, $R^I_{\ell}$, is
  plotted against $\lambda=\ell/{\C N}$ and $y$. 
  The figure does not depend on ${\C N}$. 
  Note that in calculating $R^I_{\ell}$ we are using the slow-roll approximation which is not reliable
  after $y_{end}$.
  For larger ${\C N}$, the value of $y_{end}$ tends to be smaller, and hence the domination of light
  axions tends to be stronger.}
   \label{fig3}
\end{figure}

From Fig.\ref{fig2} we can infer  that  heavy fields lose energy quite rapidly  and that  the later stage of
${\C N}$-flation is driven solely by the light fields.
To see this quantitatively let us introduce the ratio of the potential energy of $\ell$
lightest fields to that of all ${\C N}$ fields, defined by
\begin{eqnarray}
R_\ell\equiv\frac{W_{light}}{W_{total}}
=\frac{\sum_{i=1}^{\ell} V_{i}}{\sum_{i=1}^{\C N} V_{i}}\label{ratio}.
\end{eqnarray}
Using the extrapolated slow-roll solution ($I$), this ratio can be evaluated as
\begin{eqnarray}
R_\ell^I(y)
=\frac{\sum_{i=1}^{\ell}\left<y^{x_i}\right>}{\sum_{i=1}^{\C N}\left<y^{x_i}\right>}
=\frac{\left<y^x\right>\big|_1^{\tilde x_\ell}}{\left<y^x\right>}.
\label{eqn:Rell}
\end{eqnarray}
The ratio of the number of light fields to all fields is similarly
\begin{eqnarray}
\lambda\equiv\frac\ell{\C N}=\left< 1\right>\big|_{1}^{\tilde x_{\ell}}.
\end{eqnarray}
Fig.\ref{fig3} shows $R_\ell^I(y)$ versus $\lambda$ and $y$. 
Note that this plot does not depend on ${\C N}$ (however $y_{\C N}$ and $y_{end}$ do depend on ${\C N}$).
For small $y$ (i.e. at late time), $R^I_{\ell}$ approaches  $1$ for even small values of $\lambda$,
indicating that the potential energy is dominated by light axions.

Table \ref{table2} summarizes the amount of the total potential energy carried by the lightest $10\%$ 
of the axions  at 
$\sigma=\sigma(y_{end})$, for ${\C N}=100, 200, 400,1500$.
These values are calculated using both,  solution I and numerical computations. 
The results clearly show that for large ${\C N}$,  the potential energy is carried by only a small portion of lightest axion fields. 
Using (\ref{eqn:Rell}), it is not difficult to check that this tendency becomes stronger when ${\C N}$ is larger. 
This in turn justifies our usage of  the approximation given by solution $I$, which corresponds to using slow roll for all fields, even if  the slow roll condition $\eta_i<1$ is violated for the majority of fields;
all in all, heavy fields do not contribute much to the total potential energy, and henceforth inflation.

\begin{table}[htdp]
\begin{center}
\begin{tabular}{c||c|c}
${\C N}$ & solution I & Numerical \\
\hline
100 & 62.9 & 63.3 \\
200 & 78.4 & 80.0 \\
400 & 88.0 & 89.9 \\
1500 & 96.5 & 97.5
\end{tabular} 
\caption{The ratio of the potential energy carried by the lightest axions  ($10\%$) with respect to the total potential energy,
evaluated at the end of slow-roll regime $\sigma=\sigma(y_{end})$. 
The table shows both semi-analytic and numerical results.
\label{table2}}
\end{center}
\end{table}

\section{Preheating\label{icp} }

Now let us discuss the physics of preheating.
In order to solve problems of the standard big-bang cosmology the number of e-foldings 
must be large enough; with this in mind,  we assume ${\C N}=1500$ in the following. 
If we forget about its string theoretical origin (which we do in this paper), ${\C N}$ can be even larger,
giving rise to a (harmless) larger number of e-folds.
Our choice of ${\C N}$ is for definiteness, and also for numerical tractability.

\subsection{The model of preheating}

We first  recall that the axion mass scale is observationally constrained. 
The power spectrum for multi-field models is computed by the $\delta N$-formalism 
\cite{Sasaki:1995aw} and for ${\C N}$-flation it becomes \cite{Easther:2005zr,Gong:2006zp}
\begin{eqnarray}
\mathcal{P}_{\mathcal{R}}
&=&\sum_i\frac{m_i^2\varphi_i^2}{96\pi^2 M_{P}^6}\sum_j\varphi_j^2\nn\\
&=&\frac{{\C N}^2m_{0}^2\varphi_{0}^{*4}}{96\pi^2M_P^6}
\left<y^x\right>\left<x^{-1}y^x\right>,
\end{eqnarray}
where in the second line the MP distribution and equal energy initial conditions have been used.
To be specific, if we evaluate $\mathcal{P}_{\C R}$ at $t=t_*$ (corresponding to large scales) 
we obtain
\begin{eqnarray}
\mathcal{P}_{\mathcal{R}}
=\frac{\mathcal{N}^2m_{0}^2\varphi_{0}^{*4}}{96\pi^2M_{P}^6}\left<x^{-1}\right>.
\end{eqnarray}
Comparing this with 
${\C P}_{\C R}\approx 2.3\times 10^{-9}$
from WMAP measurements \cite{Bennett:2003bz,Spergel:2003cb,Peiris:2003ff},
we find the mass of the lightest field 
\begin{eqnarray}
m_{0}\approx 2.4\times10^{-6} M_{P}\,, 
\label{m0value}
\end{eqnarray}
for ${\C N}=1500$, $\beta=0.5$ and $\varphi_{0}^*=M_P$.
Consequently,  the average mass is
$
\bar{m}=m_{0}/(1-\sqrt\beta)\approx 8.1\times10^{-6} M_{P}\,.
$
In what follows, we assume these values for the mass parameters; all  other masses follow
via the MP-distribution.

In the previous section we have seen that the late stage of ${\C N}$-flation is mainly driven by a few light axions;
we naturally expect that preheating is triggered by these lightest axions, coupled to a matter field, at least before back reaction becomes important.
Thus, in this section we  focus on the relevant $\tilde{\C N}=150$ light axions, which carry more than
95 \% of the total potential energy at $y=y_{end}$.
It is important to note at this point that the relevant mass scale for
preheating is set by the mass of the light fields and not the average mass
$\bar	{m}$.
For the matter into which the inflatons decay, we consider a massless bosonic field $\chi$ 
coupled to the axions via the coupling $\half g^2\varphi_i^2\chi^2$, 
where we assume for simplicity an identical coupling constant to each $\varphi_i$. 
The model we consider is then described by the following Lagrangian,
\begin{eqnarray}
{\C L}&=&-\sum_{i=1}^{\tilde{\C N}}\left\{\half g^{\mu\nu}\nabla_\mu\varphi_i\nabla_\nu\varphi_i
+\half m_i^2\varphi_i^2+\half g^2\varphi_i^2\chi^2\right\}\nn\\
&&-\half g^{\mu\nu}\nabla_\mu\chi\nabla_\nu\chi.
\end{eqnarray}
The equations of motion are 
\begin{eqnarray}
&&\ddot\varphi_i+3H\dot\varphi_i+\left(m_i^2+g^2\left<\chi^2\right>\right)\varphi_i=0,\label{eqn:EOM1}\\
&&\ddot\chi_k+3H\dot\chi_k+\left(\frac{k^2}{a^2}+g^2\sum_i\varphi_i^2\right)\chi_k=0,\label{eqn:EOM2}\\
&&3H^2=\half\sum_i\dot\varphi_i^2+\half\sum_i m_i^2\varphi_i^2\nn\\
&&~~~~~~ +\half\left<\dot\chi^2\right>+\half g^2\left<\chi^2\right>\sum_i\varphi_i^2,\label{eqn:EOM3}
\end{eqnarray}
where $\chi_k$ is the mode operator of the matter field and $\left<\cdot\right>$ is the mode sum over 
$k$. 
We consider the axions and gravity as the background, and ignore backreaction from the matter field
$\chi_k$; consequently, $\left<\chi^2\right>$ and $\left<\dot\chi^2\right>$ are set to zero. 

\subsection{Parametric resonance in the equal-mass case\label{section:equalmass}}

Before addressing the more involved preheating scenario of
$\mathcal{N}$-flation, we discuss a model with $\tilde {\C N}=150$ inflatons having the same 
mass, $m_i\equiv m$.

In this case, the equal energy initial conditions set the same initial values for all $\tilde{\C N}$
axions, so that the evolution of the $\tilde{\C N}$ axions is identical.
Neglecting  backreaction of the matter field we can write the equations of motion as,
\begin{eqnarray}
&&\ddot\varphi_i+3H\dot\varphi_i+m^2\varphi_i=0,\nn\\
&&\ddot{\chi}_k+3H\dot{\chi}_k+\left(\frac{k^2}{a^2}+\tilde{\mathcal{N}} g^2\varphi_i^2\right)\chi_k=0\,, \nn\\
&&3H^2=\frac{\tilde{\C N}}{2}\left(\dot\varphi_i^2+m^2\varphi_i^2\right).
\end{eqnarray}
Defining $\varphi\equiv\sqrt{\tilde{\C N}}\varphi_i$, the equations of motion reduce to those of the well-understood single field model, yielding  non-perturbative preheating
\cite{Dolgov:1989us,Traschen:1990sw,Greene:1997fu,Kofman:1997yn,Bassett:2005xm}.
The Klein-Gordon equation for $\varphi$ reads
\begin{eqnarray}
\ddot{\varphi}+3H\dot{\varphi}=-m^2\varphi\,,
\end{eqnarray}
whose solution is approximated during the preheating era by
\begin{eqnarray}
\varphi(t)=\Phi(t)\sin(m t)\,, \label{varphianalytic1}
\end{eqnarray}
where $\Phi(t)=\sqrt{8}/(\sqrt{3} mt)$ \cite{Kofman:1997yn} is a slowly decaying amplitude due to 
Hubble friction. 
The corresponding equation for a Fourier mode of the matter field reads
\begin{eqnarray}
\ddot{\chi}_k+3H\dot{\chi}_k+\left(\frac{k^2}{a^2}+g^2\varphi^2\right)\chi_k=0\,, \label{EOMChi}
\end{eqnarray}
where ${\bf{p}}={\bf{k}}/a$ is the physical momentum.
Due to the oscillations of the inflaton field, the mass of the matter field becomes time dependent and resonances can occur. 
To see this, introduce 
$q=g^2\Phi^2/4m^2$, $\tau=mt$, 
$A_k=2q+k^2/m^2a^2$ and $X_k\equiv a^{3/2}\chi_k$ so that (\ref{EOMChi}) becomes
\begin{eqnarray}
\frac{d^2 X_k}{d\tau^2}+\left(A_k-2q\cos(2\tau)\right)X_k=0\,,\label{mathieu}
\end{eqnarray}
where we also neglected the term proportional to the pressure,
$-(3/4)(H^2+2\ddot a/a)$.
If we ignore the time dependence of the amplitude $\Phi$ in $q$ and of $A_k$,
Eq. (\ref{mathieu}) is the Mathieu equation. 
It is known that parametric resonance occurs for wavenumbers $k$ within resonance bands
(see \cite{mclachlan, Kofman:1997yn} for the stability/instability chart). 
This means if $k$ is within the $n$'th resonance band,  the corresponding mode increases exponentially
\begin{eqnarray}
X_k\propto e^{\mu_k^{(n)}\tau}\,,
\end{eqnarray}
where $\mu^{(n)}_k>0$ is the Floquet index \cite{mclachlan}.
Physical parameters correspond to the region $A_k\geq 2q$ and, in particular, the zero mode $k=0$ evolves along the $A_k=2q$ line from large $q$ to $q\sim 0$, as the inflaton amplitude $\Phi$  decays slowly.
As it evolves,  the system crosses resonance bands where exponential particle production takes place.
Particle production is efficient in the large $q$ ($q\gg 1$) region,  {\em broad resonance} 
(or {\em stochastic resonance} when expansion effects are included).
For small $q$ the resonance effect is limited as it is not strong enough to hold against redshifting
$\chi_k\propto a^{-3/2}$.
Then the main concern is whether it is possible to have a large enough $q$ in a given model. 
A stringent constraint comes from radiative corrections, restricting the value of the coupling to 
$g\alt 10^{-3}$ \cite{Kofman:1997yn,Zlatev:1997vd}.

\begin{figure}[tb]
  \includegraphics[scale=0.8]{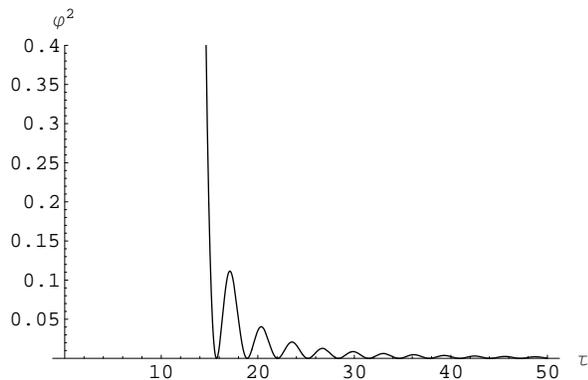}
  \caption{
  The evolution of the oscillating term $\varphi^2$ that drives  parametric resonance.
  The horizontal axis is the dimensionless time $\tau=m_0 t$.
  Since $\varphi=\sqrt{\tilde{\C N}}\varphi_i$ and $\varphi_i$ are chosen to start from $M_P=1$,
  $\varphi^2$ starts from $\tilde{\C N}=150$ at $\tau=0$. }
   \label{fig:singleBG}
\end{figure}

\begin{figure}[tb]
  \includegraphics[scale=0.8]{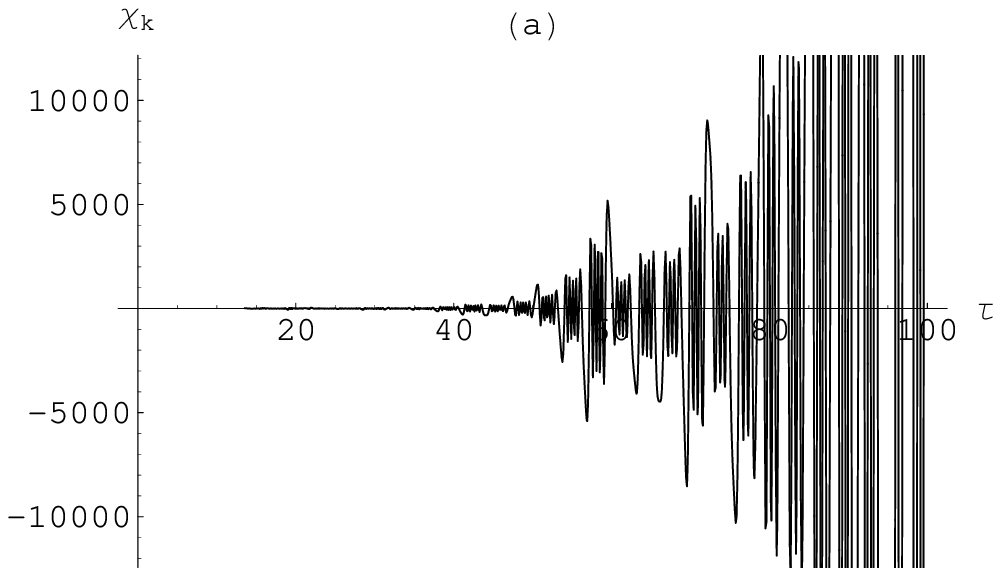}
  \includegraphics[scale=0.8]{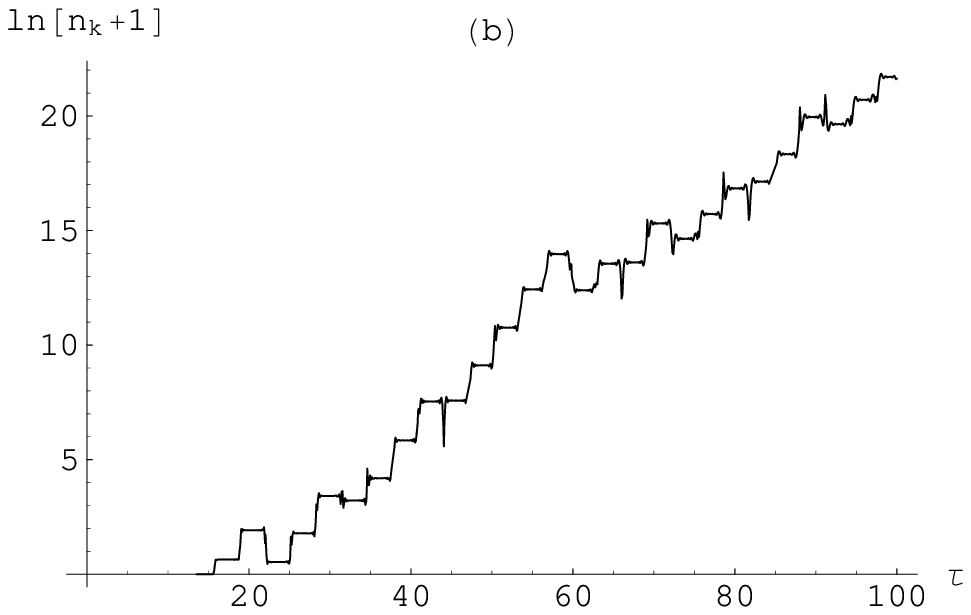}
  \caption{
  The evolution of (the real part of) the mode $\chi_k$ (a) and the occupation number $n_k$ (b).
  The initial conditions for $\chi_k$ are set by the positive-frequency solution at
  $\tau=13.5$, when the slow-roll conditions break down.
  The coupling constant is $g=10^{-3}$ and the wavenumber is chosen as $k/am = 6.0$
  at $\tau=13.5$.
  One can see amplification due to typical stochastic resonance, i.e. the overall 
  amplitude grows exponentially while there are occasional decreases of the amplitude. 
  We are ignoring backreaction so that resonances are present until $q\approx {\C O}(1)$, 
  corresponding to 
  $\tau\approx 4000$ for $g=10^{-3}$.
  Backreaction from the matter field shuts off resonances earlier. 
  }
   \label{fig:singleRH}
\end{figure}

From  the above discussion it can be deduced  that having many inflatons does not increase $q$, and  resonance 
effects are not enhanced.
Given that the equations of motion lead to the one  of the single field model, $\varphi(=\sqrt{\tilde{\C N}}\varphi_i)$ starts oscillating from the single field value $\sim 0.2 M_P$.
Each inflaton $\varphi_i$ oscillates with smaller amplitude $\Phi/\sqrt{\tilde{\C N}}$, while
$q$ is unaltered. 
In the next subsection, we compare the equal mass case to a broader mass distribution (MP mass distribution).
To this aim, we provide numerical  plots.
In Fig.\ref{fig:singleBG}, we show the evolution of the oscillating term $\varphi^2$ in the 
equal mass case,
where we have chosen $\tilde{{\C N}}=150$, $m=m_0=2.4\times 10^{-6} M_P$.
The initial values of
$\varphi_i$ are all taken to be $M_P=1$, and the initial velocities  are $\dot\varphi_i=0$. 
In Fig.\ref{fig:singleRH} the evolution of the matter field mode function $\chi_{k}$ and the comoving occupation number of particles, defined by
\beq
n_k=\half\Big(\frac{|\dot X_k|^2}{\omega_k}+\omega_k|X_k|^2\Big)-\half,
\label{eqn:nk}
\eeq
are shown for $g=10^{-3}$ and $k/a_{init}=6.0\times m$ 
(corresponding to a fastest growing mode, see \cite{Kofman:1997yn}).
Here,
\beq
\omega_k=\sqrt{\frac{k^2}{a^2}+g^2\sum_i\varphi_i^2},
\label{eqn:omega}
\eeq
which reduces to $\sqrt{k^2/a^2+g^2\varphi^2}$ in the present case.
The condition on the resonance parameter $q=g^2\Phi^2/4m^2\agt {\C O}(1)$ corresponds to
$|\varphi|^2\agt 10^{-5}$ for $g=10^{-3}$, which is roughly $\tau\alt 4000$.
We are ignoring both, backreaction and rescattering effects. 
The corresponding plots for the Mar\v cenko-Pastur distribution are shown in 
Figs. \ref{fig:MPBG} and \ref{fig:MPRH_long}  below.

\subsection{Numerical results for the Mar\v cenko-Pastur case}

Let us turn to the question of multiple fields whose masses obey the MP law.
The equations of motion of our system are (\ref{eqn:EOM1}), (\ref{eqn:EOM2}), (\ref{eqn:EOM3});
we are assuming that all inflaton fields are coupled to the same matter field with identical strength $g^2$,
and we consider only $\tilde{\mathcal{N}}=150$ axions since the heavier $90\%$ of the axions are negligible
in the later stage of ${\C N}$-flation.
We also ignore backreaction, meaning we set $\langle\chi^2\rangle=\langle\dot\chi^2\rangle=0$ 
(which, in the end, is justified since amplification of the matter field is found to be suppressed).
The initial values and the velocities of the axions at the onset of  preheating $y=y_{end}$ corresponds to  the extrapolated slow-roll solution, discussed in Section \ref{sec:bsr}:
\bea
&&\varphi_i(y_{end})=\varphi_0^*\sqrt{\frac{y_{end}^{x_i}}{x_i}},\\
&&\dot\varphi_i(y_{end})=-m_0^2\varphi_0^*\sqrt{\frac{x_iy_{end}^{x_i}}{3W_I(y_{end})}}\,.
\eea
The initial conditions for the matter field is set by the positive frequency mode funtion,
$X_k(t)=a^{3/2}\chi_k(t)\simeq e^{-i\omega_k(t-\tau_{end}/m_0)}/\sqrt{2\omega_k}$,
at $\tau=\tau_{end}=11.6$ corresponding to the onset of the preheating stage $y=y_{end}$.

\begin{description}
\item[Short time scale behavior:]

Fig.\ref{fig:MPBG} shows the evolution of $\sum_i\varphi_i^2$ until $\tau=50$.
The horizontal axis used here is the dimensionless time $\tau=m_0 t$.
Clearly, the axion masses are all different in the Mar\v cenko-Pastur case, resulting 
in a somewhat different evolution of the $\sum_i\varphi_i^2$ term from the equal-mass case. 
We can see that the oscillations are more obtuse compared to Fig.\ref{fig:singleBG}; 
this is a consequence of dephasing of the axion oscillations owed to relative mass differences.
As the time-dependent mass term oscillates, some resonance effects  for the  dynamics of 
$\chi_k$ are expected.
This is indeed the case, at least to some extent.
Fig.\ref{fig:MPRH_short} (a) shows the time evolution of the matter field mode function
($g=10^{-3}$ and $k/am_0=6.0$ at $\tau=\tau_{end}$, as in the previous section). 
The temporal enhancement of the amplitude (which is clearly seen for small $\tau$ but  becomes weaker for large $\tau$) is caused by parametric resonance with (the collective behavior of) the axions.
In contrast to the equal-mass case, the amplitude of $\chi_{k}$ decreases on average;
the resonance effect is not strong enough to resist dilution due to cosmic expansion,
even when the coupling constant is as large as $g\sim 10^{-3}$. 

\begin{figure}[tb]
  \includegraphics[scale=0.8]{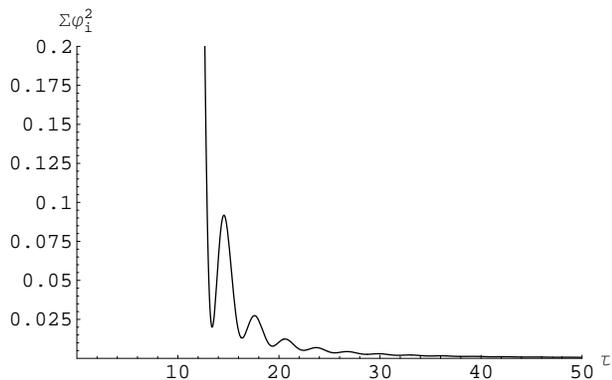}
  \caption{
  The evolution of the term $\sum_i\varphi_i^2$ that couples to the matter field.
  The time is as in Fig.\ref{fig2}.
  Oscillations are less sharp than in the equal-mass case, Fig.\ref{fig:singleBG}.
}
   \label{fig:MPBG}
\end{figure}

\begin{figure}[tb]
  \includegraphics[scale=0.8]{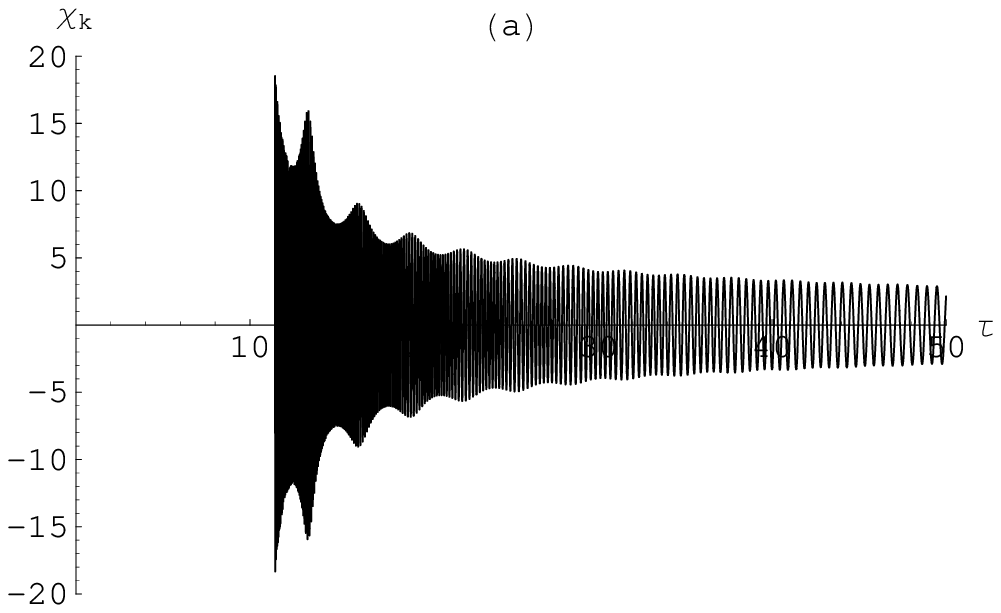}
  \includegraphics[scale=0.8]{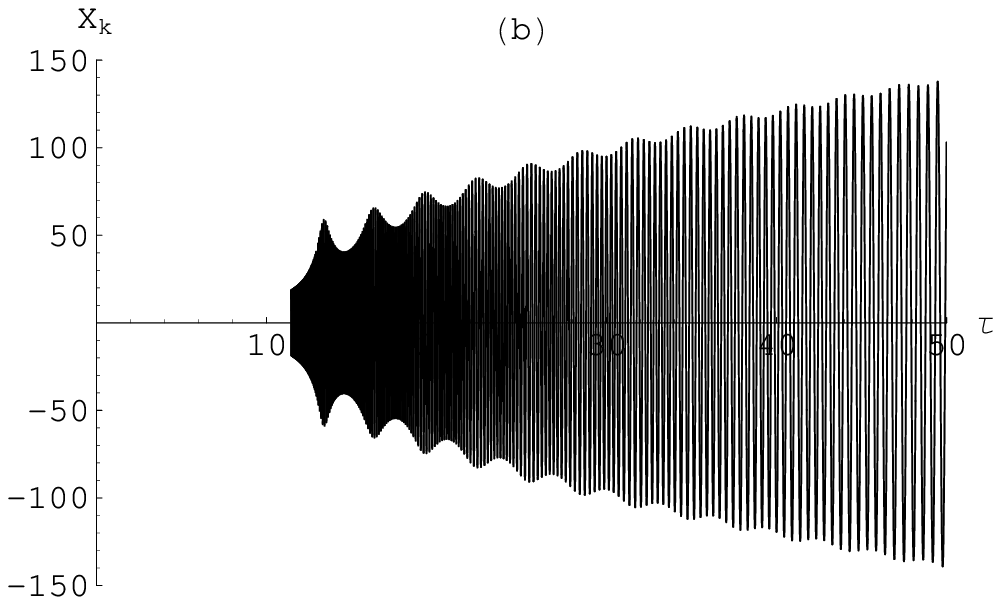}
  \caption{
  (a) The evolution of the mode function of the matter field $\chi_k$, in ${\C N}$-flation using the
  MP mass distribution.
  The coupling is $g=10^{-3}$ and the wavenumber is chosen as $k/am_0=6.0$ at 
  $\tau=\tau_{end}=11.6$.
  (b) The evolution of $X_k=a^{3/2}\chi_k$ for the same parameters.
  There are wiggles in the oscillation amplitude (these are evident for small $\tau$ and become 
  smaller for larger times) indicating some effect of parametric resonance.
  This resonance is, however, not strong enough and the amplitude of $\chi_k$ decays on 
  average.
  }
   \label{fig:MPRH_short}
\end{figure}

One can separate out the effect of cosmic expansion by looking at the comoving field
$X_k=a^{3/2}\chi_k$.
The equation of motion for $X_k$ is 
\begin{eqnarray}
\ddot X_k+\left[\frac{k^2}{a^2}+g^2\sum_i\varphi_i^2-\frac{3}{4}(2\dot H+3H^2)\right] X_k=0,
\label{eqn:XEOM}
\end{eqnarray}
where the last term in the square bracket is proportional to the pressure and is very small
during reheating.
Fig.\ref{fig:MPRH_short} (b) shows the numerical plot of the evolution of $X_{k}$. 
We can see that the peaks occur when $\sum_i\varphi_i^2$ of Fig.\ref{fig:MPBG} reaches local minima, 
and $\dot\omega_k/\omega_k^2$ becomes large (see Fig.\ref{fig:8}), i.e. when
the system becomes less adiabatic; this is characteristic of parametric resonance. 
In contrast to the equal-mass case, the minima of the mass term do not approach zero due to 
dephasing, caused by the relative mass differences of the axion fields.
This yields relatively small $\dot\omega_k/\omega_k^2$ and makes preheating inefficient.

One can see that the amplitude of $X_k$ exhibits power-law like growth on average.
This growth does not necessarily mean production of particles, since it is mainly due to
redshift \cite{Kofman:1997yn}.
When $k$ is large the power $X_k\sim a^\gamma$ tends to $\gamma\approx 0.5$, which is understood as follows:
in the mass term of (\ref{eqn:XEOM}), 
$k^2/a^2$ is dominant and it stays dominant since 
$a^{-2}\sim t^{-4/3}$, $\varphi_i^2\sim t^{-2}$, $\dot H\sim t^{-2}$ and $H^2\sim t^{-2}$
as $a\sim t^{2/3}$ during preheating.
Then (\ref{eqn:XEOM}) becomes $\ddot X_k+Ct^{-4/3}X_k=0$ for some constant $C$, which is exactly soluble in the form of $X_k\sim t^\alpha F(t)$ where $F(t)$ is a fast oscillating function; discarding the decaying solution we find $\gamma=3\alpha/2=0.5$.
For smaller $k$, $X_k$ grows faster than $\sim a^{0.5}$;
for $k\approx 0$ we find $\sim a^{0.75}$ numerically.

\begin{figure}[tb]
  \includegraphics[scale=0.8]{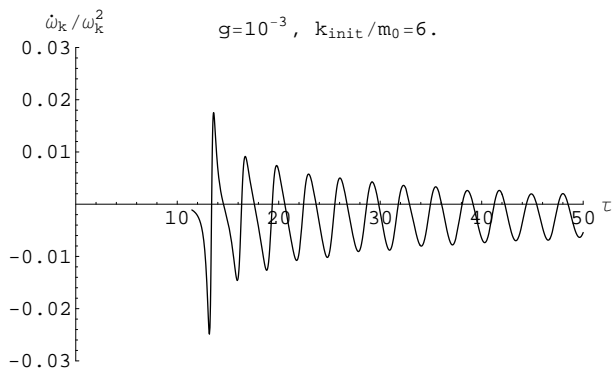}
   \caption{The adiabaticity parameter $\dot\omega_k/\omega_k^2$, for $\tau\alt 50$.
   The slight negative shift is due to the cosmic expansion. 
\label{fig:8}}
\end{figure}

%

\item[Long time scale behavior:]

in the equal-mass model (see section \ref{section:equalmass}), with a large enough value of $g$, the resonance parameter is $q$ $\gg 1$ and resonances arise for reasonably long time scales, specifically,  up until $\tau\approx 4000$ for $g=10^{-3}$ (ignoring backreaction). 
Similarly, in the MP case, even though there is no well-defined $q$-parameter, resonances can ensue during short time intervals for large $\tau$, again assuming a similar large coupling $g$. 
In this case, the collective behaviour of the axions is crucial and the adiabaticity parameter
$\dot\omega_k/\omega_k^2$ shows a rather complicated behaviour (see Fig.\ref{fig:9} (a)).
Since  the mass differences between the neighbouring axions in (\ref{eqn:MidPointMass}) 
with ${\C N}=1500$ is typically of order of $\Delta m^2\approx {\C O}(10^{-2})\times m_0^2$, once dephased, the axions' collective oscillations return to near-coherence in time scales of order
$\Delta\tau\approx {\C O}(10^2)$, causing beats in the effective mass for $\chi_k$ 
(see Fig.\ref{fig:9} (b)).
In Fig. \ref{fig:MPRH_long} we show the evolution of $\chi_k$ until $\tau=2000$ (a), 
and the evolution of the comoving occupation number $n_k$ calculated for 
$X_k$ (b). 
In this example, there is some amplification due to parametric resonances around $\tau\approx 450$.
On these time scales, for $g\agt 10^{-3}$ and small $k$, we find the occasional amplitude 
enhancement of a few orders of magnitude.
For larger wavenumbers ($k/am_0\agt 10^4$ at $\tau=\tau_{end}$) we find somewhat different
behaviour of $n_k$.
The overall amplitude of $\dot\omega_k/\omega_k^2$ becomes smaller but the spikes at large
$\tau$ remain. 
Consequently, the bursts at $\tau\approx 450$ disappear and the late time dynamics is dominated by 
a random-walk like behaviour.
These resonance effects are, however, not frequent or long enough to dominate preheating.

\end{description}

To summarize, we saw that preheating of a single matter field is not due to explosive particle production in $\mathcal{N}$-flation; even though there is some amplification, it is too weak in small time scales and not very frequent in large time scales, to compete with the dilution due to Hubble expansion.
We have also studied parameters not presented above, including larger values of the coupling $g$;
for $g=3\times 10^{-3}$ the resonance is barely sufficient to compete with the Hubble expansion.
For the MP parameter $\beta = 0.7$ and $0.9$, we found similar results (inefficient resonance).
The physical reason for the suppression of parametric resonance can be understood as follows: 
the axions are all out of phase, averaging out each other's contribution, so that the driving term 
$\propto \sum \varphi_i^2$ in the equation of motion for $\chi_k$ does not provide a coherent oscillatory 
behavior that is needed for  efficient parametric resonance. 
Hence, instead of an exponential increase, we observe power-law like behavior
$\chi_k\sim a^{\gamma-3/2}$ in time scales $\tau\alt 200$, where $\gamma$ is typically between 
$0.5$ and $0.75$ for the parameter region we have studied.
For longer time scales $\tau\approx 200\sim 2000$, we find occasional particle production, although
these are not strong enough to dominate preheating.
This conclusion differs from the common lore, namely, that parametric resonance effects are crucial for preheating \cite{Bassett:2005xm}.

\begin{figure}[tb]
  \includegraphics[scale=0.8]{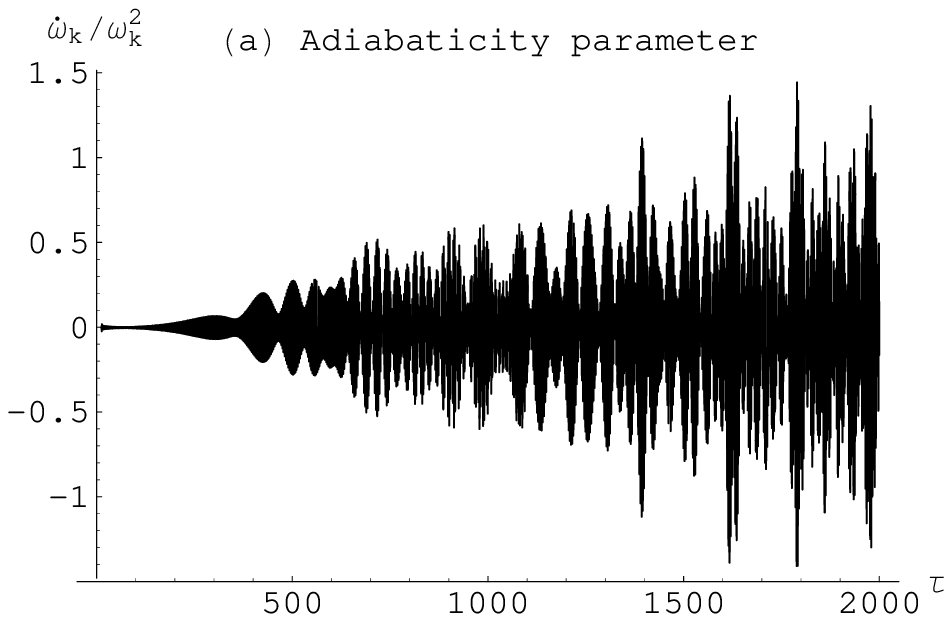}
  \includegraphics[scale=0.8]{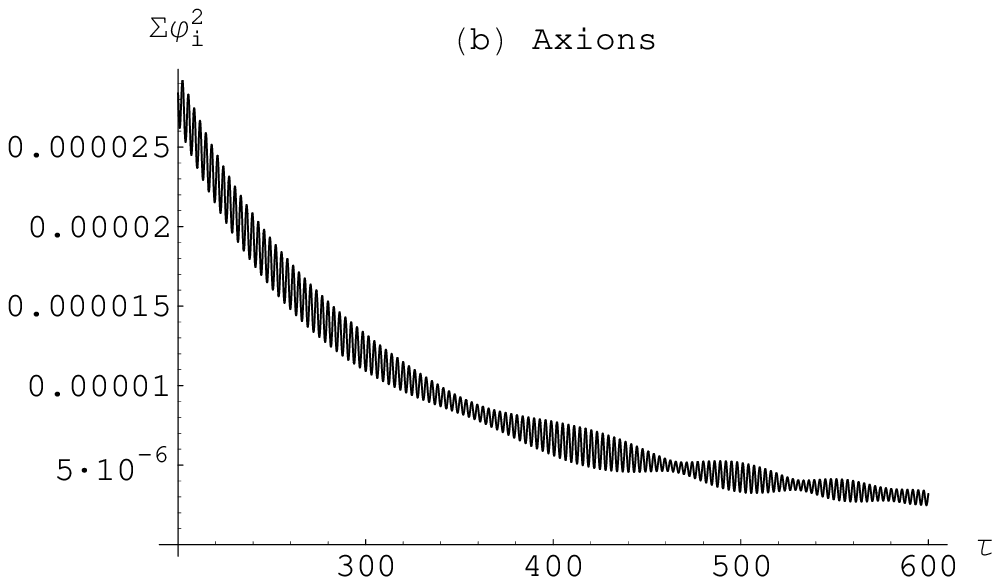}
   \caption{(a) The long time scale behaviour of the adiabaticity parameter $\dot\omega_k/\omega_k^2$.
   (b) The sum of axions' squared amplitudes $\sum_i \varphi_i^2$, for the time scale $\tau=200$ 
   to $600$. The coupling $g$ and the wavenumber $k$ are the same as in Fig.\ref{fig:MPRH_short}.
\label{fig:9}}
\end{figure}

\begin{figure}[tb]
 \includegraphics[scale=0.8]{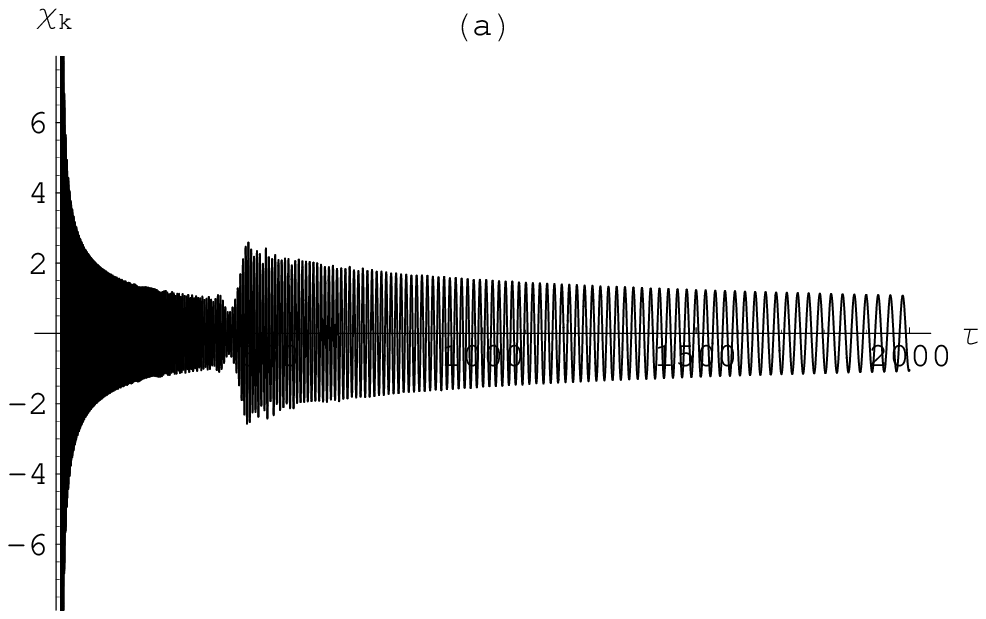}  
 \includegraphics[scale=0.8]{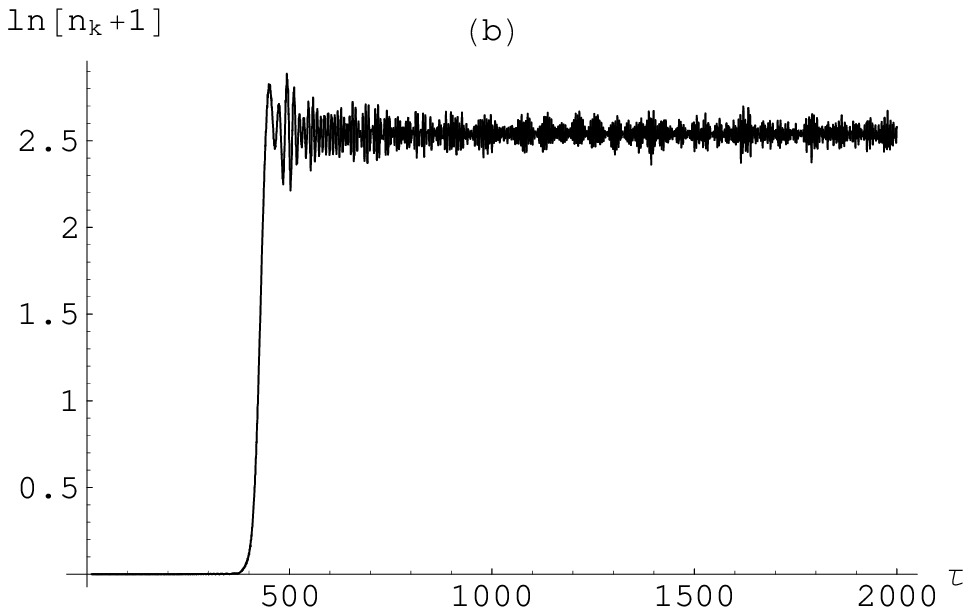}
\caption{
(a) Long time behavior of $\chi_k$, exhibiting short lived, weak resonances around $\tau\approx 450$.
The choice of parameters is the same as in Fig.\ref{fig:MPRH_short}.
(b) The comoving occupation number $n_k$ calculated for $X_k$.
\label{fig:MPRH_long}}
\end{figure}

\section{Conclusions \label{sec:conc}}

In this paper we studied the late time dynamics of ${\C N}$-flation, a string motivated realization of assisted inflation, assuming the Mar\v cenko-Pastur mass distribution (arising from random matrix theory) and equal-energy initial conditions at the onset of slow roll inflation. We provided analytic and numerical calculations of the intermediate phase after the slow roll conditions are violated for heavy fields, but before preheating commences. We find that the majority of the energy at the onset of preheating is carried by the axions with light masses, because $\sim 90 \%$ of the energy is carried by only $\sim 10 \%$ of the fields. Thus, only these light fields need to be taken into account during preheating. 
To study preheating, we coupled a single massless bosonic matter field $\chi$ to the axions $\varphi_i$, assuming the same coupling constant $g^2$ between $\chi$ and $\varphi_i$. 
Within this setup, we solved for the evolution of the matter field numerically, including the expansion of the universe, and found  power-law like behavior in short time scales and occasional, not very frequent resonance amplifications in long time scales in the parameter region that would give rise to stochastic resonance in single field models.
In particular, the growth of the matter field is generically not strong enough to resist redshifting due to cosmic expansion. 
As a result, the old theory of perturbative preheating (see e.g. \cite{Kofman:1997yn}) applies to this scenario and not parametric resonance models. 
The outcome is desirable for the model, as there is no danger of producing unwanted relics.
The prediction of this model is hence rather different from the accepted view that parametric resonance effects are crucial for preheating \cite{Bassett:2005xm}.

The analysis presented in this paper is dependent upon
several assumptions, such as the chosen matter content and coupling constants.
In particular, we considered only one matter field coupled to the whole spectrum of axion fields with the 
same coupling strength $g$. 
Although we believe this choice to be reasonable and the model to be rather generic, dropping
some of the assumptions might  change the scenario.
For instance, it is argued in \cite{Bassett:1997gb,Bassett:1998yd} that  the oscillations of multiple
inflatons (with irrational mass ratios) can enhance drastically  the decay rate  (Cantor preheating). 
This argument is based on two pillars: first, theorems in spectral theory indicate that stability bands vanish \cite{JMoser,Bassett:1997gb,Bassett:1998yd} in the case of more fields whose masses are not related by rational numbers.
Second, numerical evidence in two field models indicate a slight enhancement of particle production for well chosen parameters \cite{Bassett:1998yd}. 
In the latter study of Cantor preheating, dephasing of fields is unimportant since only two fields are considered.
Nevertheless, an enhancement effect like in Cantor preheating cannot be excluded in $\mathcal{N}$-flation, especially if only a few axions couple to a given matter field.

Besides Cantor preheating, we would like to comment on yet another effect.
It has been shown in  \cite{Zanchin:1997gf,Zanchin:1998fj} that noise on top of an oscillating driving force can also enhance resonant particle production\footnote{
However, see \cite{Ishihara:2004sx}.}. 
This phenomenon could also occur in multi-field inflation, if preheating is dominated by one or two fields: the oscillations of the many other fields would then act similar to noise, potentially enhancing preheating.

Further, we have ignored back reaction and rescattering during preheating, since modifications due to these two effects should be minor.
Explosive $\chi$-particle production due to parametric resonance is irrelevant in  $\mathcal{N}$-flation, as argued above, so that $\langle \chi^2 \rangle$ remains small;
in addition, their inclusion would only diminish resonance effects further.

\begin{acknowledgments}

We thank Thorsten Battefeld and Richard Easther for early discussions motivating this work, as well as helpful conversations with Shinta Kasuya. We also thank Kari Enqvist and Dmitry Podolsky for helpful comments.
D.B is supported by the EU FP6 Marie Curie Research and Training Network "UniverseNet" (MRTN-CT-2006-035863). S.K is supported by the Academy of Finland Finnish-Japanese Core Programme grant 112420. 
Some of the numerical calculations were carried out using computing facilities at the
Yukawa Institute, Kyoto University.

\end{acknowledgments}

\appendix


\appendix
\section{Further analytic estimates \label{AppA}}

In section~\ref{sec:A} we developed a semi-analytic solution ($I$) which underestimates
the numerical values of the potential energy $W$ (see Table \ref{table1}).
Here we present a second analytic approximation, which we  call solution ($II$), that gives
an upper bound for $W$ after the slow roll condition for the heaviest field is violated, to check the numerical solution in Table \ref{table1}.
In addition, both analytic solutions can be used for arbitrarily large numbers of fields 
that would not be tractable via a brute force numerical integration.

The basic idea consists of holding fixed heavy fields as soon as the corresponding $\eta$ becomes of order one:
first, we take  the continuum limit so that we  can make  use of the
Mar\v{c}enko-Pastur law for the continuous mass variable $1\leq x \leq
\xi$. Second, we  partition this interval into $\mathcal{M}$ bins
according to a  simple rule and denote the upper boundaries of
bins with $X_A$, $A=1,\dots ,\mathcal{M}$, so that
$X_{\mathcal{M}}=\xi$.
Third, whenever $\eta_A$ (corresponding to some
$X_A$) becomes of order one, we  hold fixed all fields with masses in
the $A$'th bin. Naturally, one recovers the full microscopic model if
one takes $\mathcal{M}=\mathcal{N}$ and uses the Mar\v{c}enko-Pastur
law as a rule for choosing the bins
so that $X_A=\tilde m_A^2/m_0^2$.

Taking $\mathcal{M}<\mathcal{N}$ leads to a coarse-grained model which is more tractable, 
but one pays the price of having a larger $W$. 
This  approximation is justified as long as the energy left in the heavy fields is small
compared to the energy in the light fields; 
$\mathcal{M} \sim 50$ suffices for the range of $y$-values that we are interested in\footnote{
Note, generically $\sigma_I(y) \neq \sigma_{II}(y)$ because firstly, a number of fields
are artificially held fixed and no longer contribute to the path
length $\sigma$, and secondly, the total potential energy is bigger so that
light fields evolve slightly slower. 
Only small corrections result, since fixed 
fields are already near the minimum of their
potential, not contributing much to
$\sigma$ anyhow. Moreover, we demand $R<1$. Consequently, we have $y_{II}\sim y_I$ and $\sigma_{II}\sim\sigma_I$.
}.

We now proceed  to compute $W_{II}$, $\sigma_{II}$ and $N_{II}$ as outlined
above. We  assume first a partition $\{X_1,\dots,X_{\mathcal{M}}\}$ of the interval $1\leq x \leq \xi$ and denote with $Y_A$ the values of $y$ where $\eta_{(\mathcal{M}-A+1)}=1$ (note that $Y_A<Y_B$ if $A>B$). If we further denote the energy $W_{II}$ that is valid in the range $Y_A<y<Y_{A-1}$ with $W_A$, we can calculate the corresponding $Y_A$ as the solution to
\begin{eqnarray}
W_A(Y_A)=m_0^2 X_{\mathcal{M}-A+1} \,,
\end{eqnarray} 
starting with $W_1\equiv W_{I}$. Note that $Y_1=y_\C{N}$ from (\ref{yN}), as it should. We can then compute $W_A$ for $A\geq 2$ to
\begin{eqnarray}
&&W_{A}(y)\\
&&~=\frac{\mathcal{N}}{2}m_0^2\varphi_0^{*2}
\left(\left<y^x\right>\Big|_1^{X_{\mathcal{M}-A+1}}+\sum_{n=1}^{A-1}\left<Y_n^x\right>
\Big|_{X_{{\C M}-n}}^{X_{{\C M}-n+1}}\right)\,.\nn
\end{eqnarray}
Similarly, if we denote with $\sigma_A$ the effective field which is valid in the range $Y_A<y<Y_{A-1}$ (so that $\sigma_1=\sigma_I$), we arrive at
\bea
\nonumber \sigma_A(y)&=&\sigma_{A-1}(Y_{A-1})\nn\\
&&+\frac{\sqrt{\mathcal{N}}}{2}\varphi_0^{*}\int_y^{Y_{A-1}}
\frac{1}{s}\sqrt{\left<x s^x\right>\Big|_1^{X_{M-A+1}}}ds\,,
\eea
and finally the number of e-folds becomes (with $Y_0= 1$)
\begin{eqnarray}
N_{A}(y)=N_{A-1}(Y_{A-1})+\int_{y}^{Y_{A-1}}\frac{W_A(s)}{2m_0^2s}ds\,.
\end{eqnarray}

In an appropriate large $\mathcal{M}$-limit the above approximation becomes independent of
the partition, which is of course our aim. We would like to use
$W_{II}$ up to when  $W_I$ is no longer a viable lower bound for the
true energy $W$, that is, until $\sigma\approx
\sigma_I(y_{end})$. This is possible by tuning $X_1$ such that $\sigma_{II}(Y_{\mathcal{M}})\approx \sigma_{I}(y_{end})$ \footnote{
We choose $X_1$ as large as possible so that $R<1$, while keeping $X_1$ small enough to ensure that the solution ($II$) remains applicable up until $\sigma_I$ leaves slow roll; 
thus we demand $\sigma_{II}(Y_{\mathcal{M}})\approx\sigma_{I}(y_{end})$, 
leading to $X_1=1.75$ for $\mathcal{N}=1500$.
Simultaneously, to distribute the remaining bins, we choose $(\mathcal{M}-2)/2$ narrow bins from 
$X_1$ to $X_{\mathcal{M}/2}\approx 11$ (the MP distribution peaks in that region), followed by larger bins up until $X_{\mathcal{M}}=\xi\approx 34$. For $\mathcal{N}=1500$ even $\mathcal{M}\alt 50$ yield results insensitive to the chosen partition.}.

That way, the energy ratio of heavy to light fields becomes
\begin{eqnarray}
R&\equiv&\frac{W_{heavy}}{W_{light}}
=\frac{\frac{2}{\mathcal{N}m_0^2}W_{II}(Y_{\mathcal{M}})-\left<Y_{\mathcal{M}}^x\right>\Big|_1^{X_1}}{\left<Y_{\mathcal{M}}^x\right>\Big|_1^{X_1}}
\end{eqnarray}
which  has to be smaller than one (see Table \ref{table3}), so that we can trust our approximation.

\begin{table}[htdp]
\begin{center}\begin{tabular}{c||c|c|c|c|c|c}
${\C N}$&$X_1$&
$\sigma_{II}(Y_{\C M})$ & $W_{II}(Y_{\C M})$&$\frac{W_{II}(Y_{\C M})}{W_{I}(y_{end})}$ & $\frac{N_{II}(Y_{\C M})}{N_{I}(y_{end})}$&$R$\\
&&&&&& \vspace{-0.3cm}\\
\hline
200&2.90&5.40 & 2.91& 1.47&1.07 &0.97\\
400  &2.15& 8.32&2.15& 1.24& 1.01&  0.74\\
1500&1.75& 17.57& 1.75& 1.17& 1.00&0.53\\
\end{tabular} 
\caption{Semi-analytic solution $II$ for ${\C N}=200, 400, 1500$, to be compared with 
the slow roll result $I$ from table {\ref{table1}}.
Here, $R$ is the ratio between the potential energy of the heavy 
(held-fixed) fields to that of the light (dynamical) ones.
We use a mass splitting into ${\C M}=50$ bins.
$X_1$ is chosen such that $\sigma_{II}(Y_{\mathcal{M}})\approx\sigma_{I}(y_{end})$. 
The approximation ($II$) approaches the slow roll result for increasing ${\C N}$. 
The numerical results in Table \ref{table1} lie nestled between the two approximations. 
\label{table3}}
\end{center}
\end{table}

We compare  ($I$) and ($II$) solutions in Table \ref{table3}, where we also vary the number of fields. 
The solutions approach each other in the large $\mathcal{N}$ limit.
Henceforth, the numerical solution is well approximated by either one in the case of $\mathcal{N}$-flation, where we deal with thousands of fields, and consequently, we are justified to use the slow roll approximation to set the initial stage for preheating.

\newpage

\end{document}